\magnification=1200 
\newskip\smallskipamount
  \smallskipamount=6truept plus2truept minus2truept
\newskip\medskipamount
  \medskipamount=12truept plus4truept minus4truept
\newskip\bigskipamount
  \bigskipamount=24truept plus8truept minus8truept
\newskip\normalbaselineskip
  \normalbaselineskip=24truept
\newskip\normallineskip
  \normallineskip=2truept
\newdimen\normallineskiplimit
  \normallineskiplimit=0truept
\newcount\interdisplaylinepenalty
  \interdisplaylinepenalty=100
\baselineskip=24truept
\lineskip=2truept
\lineskiplimit=0truept

\input epsf 
\def\omm{\Omega_m}
\def\omt{\Omega_T}
\def\omw{\Omega_w}
\def\wf{w_T} 
\def\bu{$\bullet$\quad} 
\def\bip{\bigskip}
\def\mip{\medskip} 
\voffset0.5truein 
\bip 

\centerline{\bf Understanding the Optimal Redshift Range for the 
Supernovae Hubble Diagram} 
\bip
\centerline{Eric V.~Linder} 
\centerline{Berkeley Lab}
\bip\bip 
\centerline{ABSTRACT}
\mip 
The supernovae Hubble diagram traces the expansion history of the 
universe, including the influence of dark energy.  Its use to probe the 
cosmological model can fruitfully be guided by heuristic study of the 
features of the model curves.  By relating these directly to the physics 
of the expansion we can understand simply the optimal redshift range for 
supernovae observations, requiring a survey depth of $z=1.5-2$ to distinguish 
evolution in dark energy properties. 
\bip 
\leftline{\bf Energy Densities and Acceleration}
\mip 
The Friedmann equations describe the physics of the expansion in terms of 
the energy density in the components, $\{\Omega_i\}$, and their 
equations of state $\{w_i\equiv p_i/\rho_i\}$: 
$$
\eqalignno{(\dot a/a)^2&=H_0^2\,\left[\sum_i \Omega'_i(z)+(1-\omt)(1+z)^2
\right]&(1)\cr 
\ddot a/a&=-(1/2)H_0^2\,\sum_i (1+3w_i)\Omega'_i(z),&(2)\cr}
$$ 
with $\omt=\sum_i\Omega_i=\sum_i\Omega'_i(0)$ and the energy densities of 
independent components evolving according to 
$$
\eqalignno{\Omega'_i(z)\equiv(8\pi/3H_0^2)\rho_i(z)&=\Omega_i\,(1+z)^3 
e^{3\int_0^{\ln(1+z)} d\ln (1+z')\,w(z')},
&(3)\cr 
&\to \Omega_i\,(1+z)^{3(1+w)}\qquad (w\ {\rm constant}).&(4)\cr}
$$ 

Because of these differing evolutions with redshift, the ratio of energy 
densities changes with time and the dominant dynamic component at one 
epoch, that driving the expansion rate, may give way to another component. 
Similarly the acceleration behavior of the expansion may alter, with a 
different component dominating.  
The global expansion may even change from deceleration ($\ddot a<0$) 
to acceleration ($\ddot a>0$), signifying a total equation of state 
sufficiently negative to break the strong energy condition ($w_T<-1/3$). 
We denote such switching epochs as $z_{eq}$ and $z_{ac}$ respectively. 

The redshift of equality between two components, taken here to be 
nonrelativistic matter ($w=0$) and some other component, presumably 
with negative equation of state, is when $\omm'(z_{eq})=\omw'(z_{eq})$: 
$$
z_{eq}=\left(\omw\over\omm\right)^{1\over -3w}-1,\eqno(5)
$$ 
when $w$ is constant (note $\Omega$'s are always present values, unlike 
$\Omega'(z)$). 

The switch to an accelerating universe takes place when $\ddot a=0$, at, 
assuming only two components are dynamically important, 
$$
z_{ac}=\left[-(1+3w){\omw\over\omm}\right]^{1\over -3w}-1,\eqno(6)
$$ 
again assuming $w$ constant.  The quantities $z_{eq}$ and $z_{ac}$ 
are plotted in Figure 1 for two flat models, $\omm+\omw=1$.  The results 
do not greatly change if we allow varying $w$ or nonflat models. 

\mip 
\midinsert 
\centerline{
\epsfxsize=3.5truein
\epsfbox{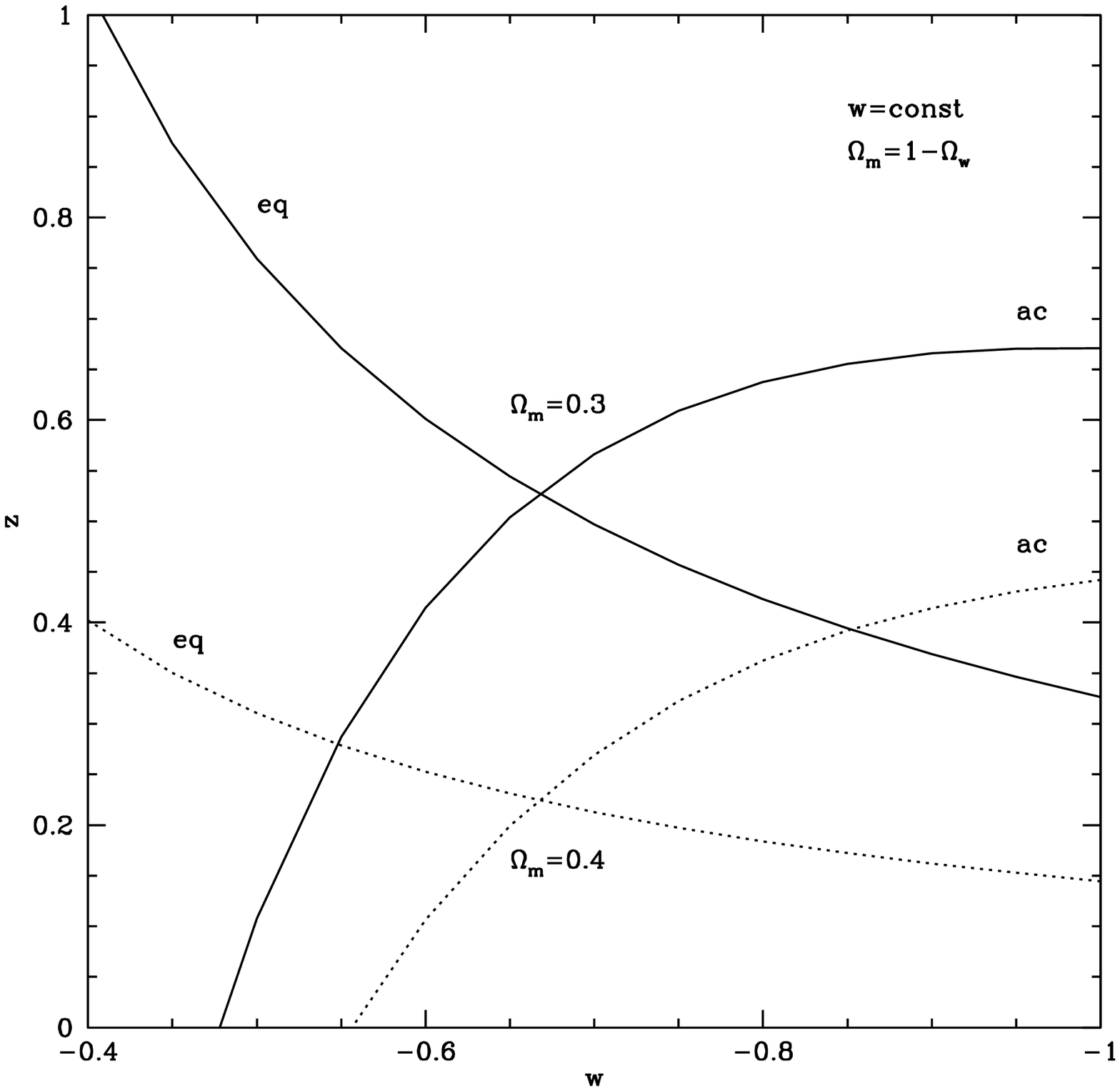}
}
\mip
\nobreak\noindent 
FIG.~1.\quad The redshifts of matter-dark energy equality and of the 
transition from decelerating to accelerating expansion are plotted 
vs.~equation of state of the dark energy, for a flat universe.  The solid 
curves have $\omm=0.3$, the dotted $\omm=0.4$. 
\endinsert 
\mip 

Most of the action in the changeover of the dynamics, both for the 
dominant energy density component and the expansion acceleration, 
happens at fairly low redshifts, $z\approx0.5$, for reasonable values 
of the matter density.  This is in marked contrast to distinguishing 
between cosmological models with fixed equation of state, say between 
models with different matter density, where the expansion behavior 
diverges at high redshifts.  Does this mean that to detect the main 
influence of a dark energy component with $w<-1/3$ we need only probe 
the expansion history to this redshift?  The answer is no, as would 
be seen from Monte Carlo simulations of the cosmological parameter 
likelihood determinations, but here we shall come to this conclusion 
using a more transparent, intuitive examination of the physics behind 
the expansion history. 

Also note that a cosmological model can have $z_{eq}<z_{ac}$ or 
$z_{eq}>z_{ac}$, i.e.~first start accelerating then become $w$ 
dominated or v.v., depending on the equation of state.  So the equation 
of state parameter, of the dark energy or the effective total energy, 
is a key variable and one that can be probed sensitively through the 
supernovae Hubble diagram out to moderate redshifts. 

In the following sections we examine the redshift range for determination 
of the equation of state, discussing use of the effective total equation 
of state, evolving $w(z)$ and ``inertia'' or ``memory'' in the 
magnitude-redshift diagram, and discrimination between models including 
maximum deviation and turnover rate characteristics. 

\bip 
\leftline{\bf Hubble Diagram} 
\mip 
The two quantities entering the Hubble diagram -- magnitude and redshift -- 
are direct tracers for the expansion history of the universe.  The magnitude 
is a logarithmic measure of the luminosity distance, related to the 
lookback time -- dimmer means further in the past, while the redshift is 
exactly the relative expansion since that time; thus $m(z)$ is $a(t)$ in 
a transparent sense.  Curvature in the diagram probes evolution in the 
expansion rate: acceleration, directly dependent on the equation of state 
as seen from eq.~(2). 

Explicitly, the relations are 
$$
\eqalignno{m(z)&\sim 5\log r_l(z),&(7)\cr 
r_l(z)&=(1+z)(1-\omt)^{-1/2}\sinh\left[(1-\omt)^{1/2}\int_0^z 
dz'\,[H(z')/H_0]^{-1}\right]&(8)\cr 
&\to (1+z)\int_0^z dz'\,[H(z')/H_0]^{-1},\cr 
H(z)/H_0&=[\omm'(z)+\omw'(z)+(1-\omt)(1+z)^2]^{1/2}&(9)\cr 
&\to (1+z)^{3/2}[\omm+\omw e^{3\int_0^{\ln(1+z)} 
d\ln(1+z')\,w(z')}]^{1/2},\cr}
$$ 
where the restricted cases are for flat universes, $\omt=\omm+\omw=1$. 

At the lowest redshifts, the luminosity distance depends on parameters as 
$$
r_l=z+{1\over2}(1-q_0)z^2+z^3 f(\omt,\langle w\rangle,\langle 
w^2\rangle),\eqno(10)
$$ 
where angle brackets denote density weighted averaging, $\langle w\rangle= 
\sum_i w_i\Omega_i/\omt$; note $\langle w\rangle=w_T$, the 
effective equation of state of all the components combined.  

The magnitude-redshift test therefore probes not only the present 
acceleration, $q_0$, but differing 
combinations of cosmological parameters in different redshift ranges. 
This is one of its great strengths in that not only is it most 
sensitive to changes, e.g.~acceleration, in the expansion 
history of the universe at recent times, but it provides 1) complementarity 
to high redshift tests, other medium redshift tests, and even itself 
over different redshift ranges; 
2) detection of evolution in the state and nature of the universe by 
probing different epochs than the cosmic microwave background, say, 
and different energy densities (smooth components) than large scale 
structure tests; 3) protection against secular or differently evolving 
systematic effects such as grey dust. 

To examine its ability to discriminate between cosmological models 
first consider models with a single component but different values 
for the magnitude of the energy density, e.g.~different $\omm$. 
The difference in magnitudes between such models is monotonically 
increasing with redshift, until $r_l\sim\omm^{-1}z$ at asymptotically high 
redshifts, so discrimination improves with ever increasing supernovae 
redshift (observational difficulties aside). 

But for a competing component, especially one with negative pressure, 
its influence is seen over a limited redshift range and little advantage 
accrues to a deep survey -- discrimination is squeezed in between the 
common linear distance-redshift law at small redshifts and the matter 
dominated relation at high redshifts.  The question addressed here is 
whether the optimal redshift range is near $z=0.5$ or higher, and how broad 
the window is. 

\bip 
\leftline{\bf Effective Equation of State} 
\mip 
First consider a series of flat, constant $w$ models with fixed 
$\omw=1-\omm$.  One might expect that their magnitude differentials 
$\Delta m_{w'}=m_w-m_{w'}$ also 
diverge with increasing redshift, and that the optimal depth is 
therefore pushed (at least formally) to high redshift.  But this is 
not quite true as seen in Figure 2.  

\mip 
\midinsert 
\centerline{
\epsfxsize=3.5truein
\epsfbox{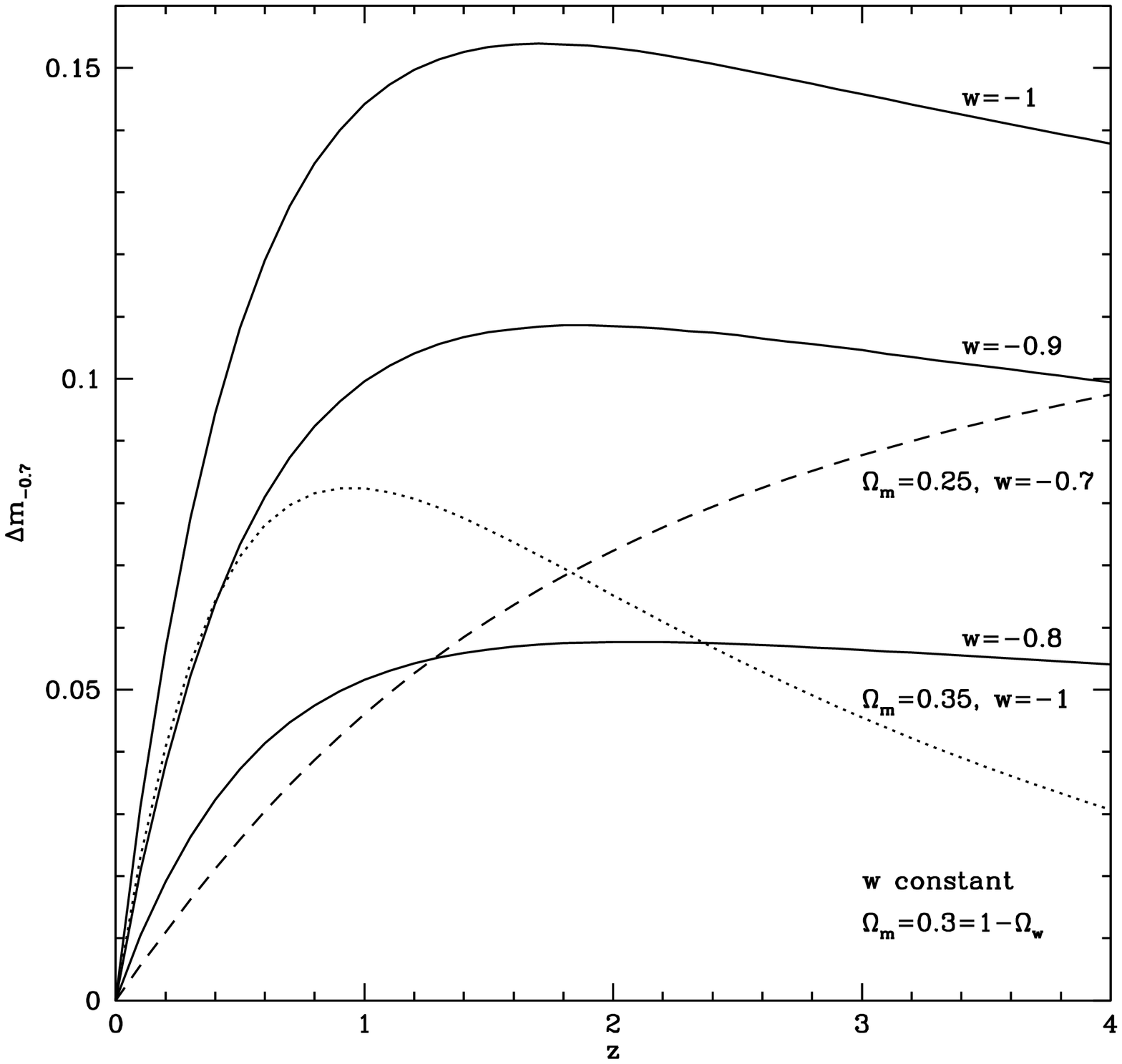}
}
\mip\nobreak\noindent 
FIG.~2.\quad The differential magnitude-redshift relation is plotted 
for various equations of state relative to the $w=-0.7$ case.  The solid 
curves have constant $w$ 
(and fixed $\omw$) but do not diverge with redshift the way curves of 
$\omw$ (at fixed $w$) would (compare the dashed curve), due to dilution 
by a decelerating component.  For comparison the dotted curve shows the 
turnover behavior upon varying both $w$ and $\omw$. 
\endinsert 
\mip 

Models with fixed equation of state but different proportions of dark 
energy and matter indeed diverge, showing increasing magnitude differences 
with redshift as illustrated by the dashed curve.  But models with fixed 
energy densities and differing equations of state (constant in time) 
level off and no great advantage accrues to probing the cosmology with 
supernovae at higher redshift.  A clear signature of models with both 
differing dark energy density and equation of state is the nonmonotonic 
behavior of the dotted curve, also a property of evolving equations of 
state as we will discuss.  These different characteristics of the shape, 
or curvature, of the magnitude-redshift relation thus provide important 
clues to the cosmological model. 

The curvature in the differential 
Hubble diagram $\Delta m-z$ probes the differential acceleration 
between the models, which receives contributions from both the negative 
pressure component and the increasingly dominant matter.  While this 
relation between curvature and acceleration is obvious at low redshift, 
as seen in equation (10), the physical correspondence continues at 
higher redshifts as shown by the following argument. 

Consider $r_l$; the key cosmological variable determining $r_l$ is 
$H^{-1}(z)$.  (It is actually the integral over redshift that enters, which 
will lead to an ``inertia'' effect discussed later.)  Whether models 
continue to diverge or turn around depends on the evolution of $H^{-1}$. 
The ``scale height'' of this evolution is 
$$
{d\ln H^{-1}\over d\ln(1+z)}=-{dH^{-1}\over dt}=-(1+q),\eqno(11)
$$ 
so the evolution in the differential Hubble diagram is indeed due to 
the differential acceleration of the models.  This in turn can be related 
to the effective, or total equation of state by $\Delta q\propto 
\Delta\wf$ since 
$$
q(z)=(1/2)\sum_i (1+3w_i)\Omega'_i(z)\bigg/\left[\sum_i \Omega'_i(z)+
(1-\omt) (1+z)^2\right],\eqno(12)
$$ 
and the total equation of state 
$$
\eqalign{\wf&\equiv p_T/\rho_T= 
\sum_i w_i\Omega'_i(z)\big/\sum_i \Omega'_i(z)\cr 
&\to w\omw'(z)/[\omm'(z)+\omw'(z)]=w\bigg/\left[1+{\omm\over\omw}e^{-3\int 
d\ln(1+z)\,w}\right],\cr}\eqno(13)
$$ 
with the restriction for the matter plus dark energy model and $w_T(0)= 
w\omw/\omt$.

Thus the key observational characteristic of curvature in the Hubble 
diagram is a direct probe of the total equation of state of the universe. 
Note that even for constant $w$ the presence of a matter component as 
well ensures 
that $\wf$ evolves with redshift, with the negative pressure gradually 
being diluted to zero at high redshift.  From eq.~(13) the dilution is 
obviously enhanced for larger $\omm/\omw$.  An interesting property is that 
the dilution can be nonlinear, since initially more negative equations of 
state are diluted more rapidly due to their diminishing energy density at 
higher redshifts, and so curves of $\wf$ can cross: $\wf(w)>\wf(w')$ 
at high redshift despite $w<w'$. 

\bip 
\leftline{\bf Evolving $w$ models} 
\mip 
The same expression, eq.~(13), holds for defining $\wf$ in the case of 
evolving dark energy.  Since $\wf$, through differential acceleration, 
forms the physical basis for 
understanding the deviation and curvature in the magnitude-redshift 
diagram, figures like Figure 3 are invaluable for an intuitive grasp 
of how models behave and how cosmological parameters can be distinguished 
as a function of survey depth. 

\mip 
\midinsert 
\centerline{
\epsfxsize=3.5truein
\epsfbox{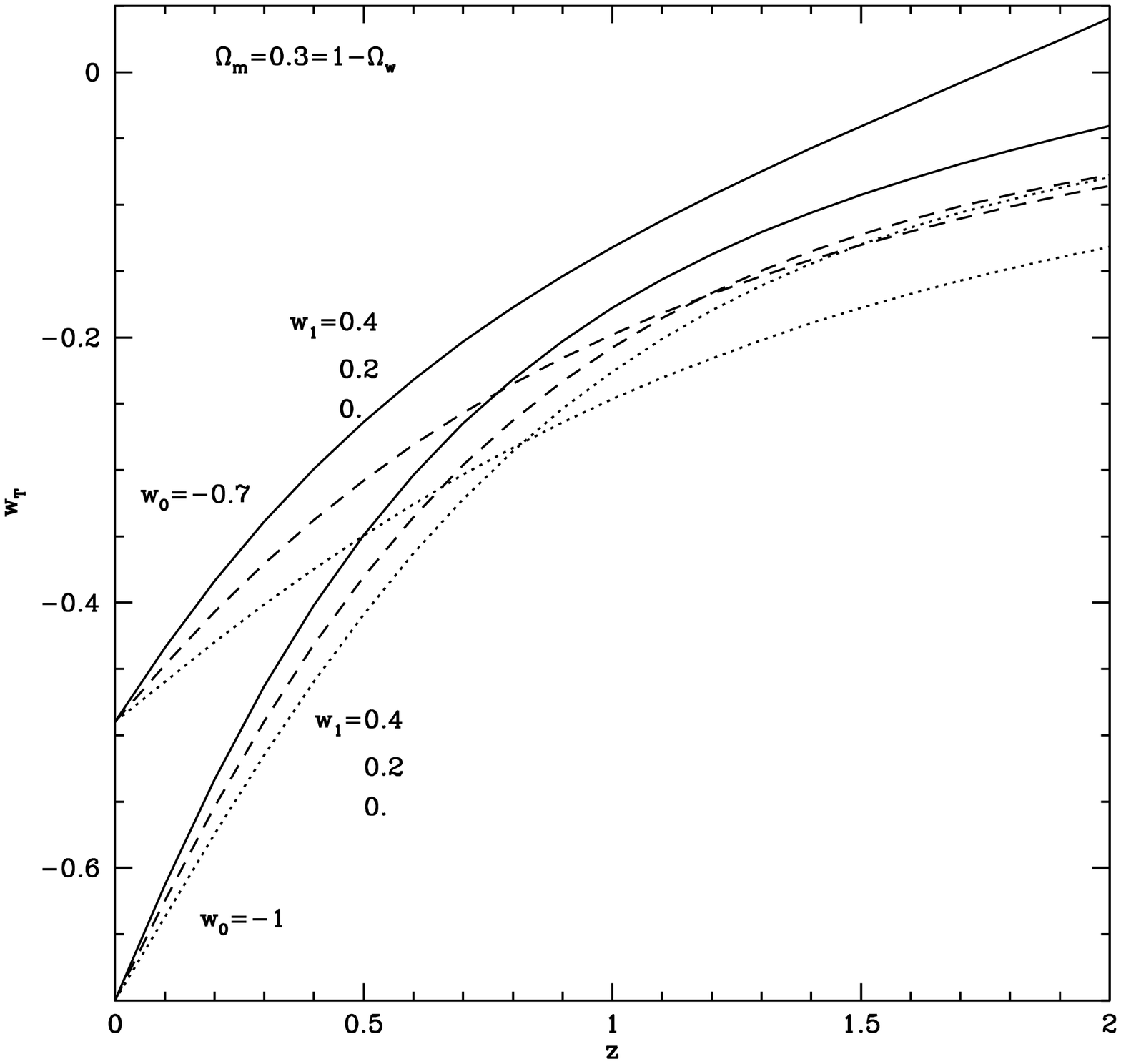}
}
\mip\nobreak\noindent 
FIG.~3.\quad The total or effective equation of state is plotted 
vs.~redshift.  Dotted curves are constant $w$ models, others have 
evolving $w=w_0+w_1z$, with $w_1=0.2$ for dashed curves, $w_1=0.4$ 
for solid curves.  All models are flat.  Even constant $w$ models 
effectively evolve with redshift due to matter dilution, approaching 
$w=0$.  Dilution is nonlinear with a model containing a more negative 
pressure component overtaking a less negative one and becoming more 
matterlike (decelerating). 
\endinsert 
\mip 

Here we see that constant $w$ models are diluted and do indeed cross, 
with the effective equation of state of the cosmological constant 
($w=-1$) model actually greater (more matterlike and decelerating of 
the expansion) than the $w=-0.7$ 
model at $z\ge 0.8$.  Evolving models are linearly parametrized, 
$w=w_0+w_1z$.  Positive $w_1$ of course increases the dilution, and 
if sustained for long enough can drive $w$ and $\wf$ positive, but one 
must be careful not to extend the linear evolution approximation 
beyond its validity. 

Using the $\wf$'s of two models as guides to their differential 
acceleration, one can deduce the behavior of their differential 
magnitude-redshift behavior $\Delta m(z)$.  One expects that 
a model with $\wf$ less than the fiducial model (hence greater acceleration 
of the expansion) would have a positive magnitude offset at a given 
redshift, i.e.~possess apparently dimmer supernovae.  This offset would 
increase with redshift until the greater dilution of $\wf$ for the 
more negative $w$ model causes it to approach the $\wf$ value of the 
fiducial.  On the magnitude diagram this would appear as a slowing of 
the deviation and eventually a turnover at approximately the redshift 
when the two $\wf$'s agree (but see the Inertia section later).  If 
the effective equations of state then keep pace, the turnover is just 
a leveling off, while if the $\wf$ curves cross then the magnitude 
deviation approaches zero (and can go negative though eventually it 
levels off at redshifts fully matter dominated). 

For example, from Figure 3 we would predict a priori that models 
($w_0,w_1$) should have the following behaviors in their 
(differential) Hubble diagrams: 

\bu (-0.7,0.4)-(-0.7,0): increasingly negative magnitude deviation with 
no turnover 

\qquad or slowing. 

\bu (-0.7,0.2)-(-0.7,0): increasingly negative magnitude deviation with 
no turnover 

\qquad but gradual slowing. 

\bu (-1,0.2)-(-1,0): increasingly negative magnitude deviation with no 
turnover 

\qquad but leveling off. 

\bu (-0.7,0.2)-(-1,0): increasingly negative magnitude deviation with 
leveling 

\qquad beyond $z=1.5$ and very gradual turnover. 

\bu -[(-1,0.4)-(-0.7,0)]: increasingly negative magnitude deviation with 
leveling 

\qquad beyond $z=0.5$ and turnover. 

\noindent The corresponding Hubble diagrams are shown in Figure 4, 
bearing out these heuristic predictions. 

\midinsert 
\centerline{
\epsfxsize=3.5truein
\epsfbox{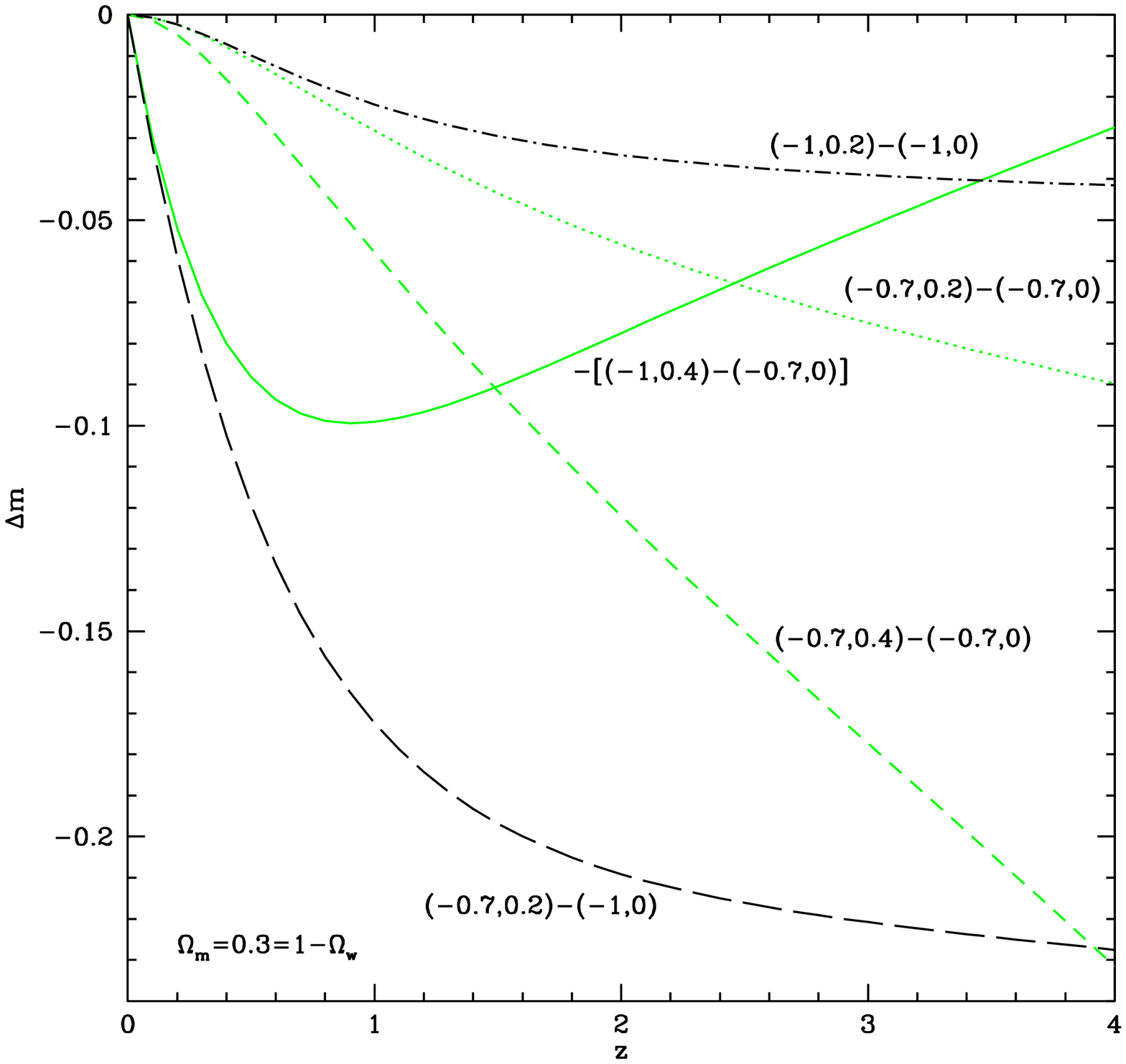}
}
\mip\nobreak\noindent 
FIG.~4.\quad Differential Hubble diagrams plot the magnitude differences 
between the $w(z)=w_0+w_1z$ models labeled with ($w_0,w_1$), exhibiting 
a variety of divergence, 
leveling, and turnover behaviors.  Heuristic analysis by means of the 
total equations of state is successful in predicting the qualitative 
features. 
\endinsert 

Thus knowledge of the equation of state $w$ for a model enables direct 
qualitative construction 
of its magnitude-redshift relation through $\wf$, including 
the sign of the deviation, the presence of leveling or turnover, and the 
sharpness of the evolution.  Quantitative results such as the size of 
the maximum magnitude deviation must be gotten by numerical evaluation of 
the luminosity distance from eq.~(8).  However we can obtain a rough 
estimate of the turnover redshift, and hence the required survey depth, 
by examining the $\wf$ curves as above.  
This method becomes more definite in the Inertia section.  

The inverse 
process of looking at a differential Hubble diagram and intuitively 
extracting the model equation of state is more problematic.  One could 
probably get a rough sense of $\wf$ at the redshift of any turnover, and 
an indication of how rapidly it was increasing at redshifts lower and 
higher than this, but more quantitative estimates necessitate numerical 
analysis.  The main virtue of the heuristic method though is estimation 
of the redshift range required to distinguish models, to find the ``sweet 
spot'' of the Hubble diagram that discriminates between different constant 
$w$ models and between constant and evolving $w$ models.  The key to this 
is the turnover from $\wf$ representing the differential acceleration. 

\bip
\leftline{\bf Inertia} 
\mip 
While the criterion on $\wf$ for the turnover works reasonably and was 
justified physically through the differential acceleration and scale length 
argument of the Effective Equation of State section, it is not the 
instantaneous 
evolution of the Hubble parameter $H(z)$ that governs the distance and 
magnitude, but rather its behavior over the entire redshift range between 
the source and observer.  This memory (or foreshadowing) of $w(<z)$ produces 
an inertia in the magnitude-redshift relation at $z$ so that it does not 
respond immediately to $\wf(z)$.  

For example, even though $\wf$ for two models may have converged, the 
inertia of the integral over $z$ keeps them from having a purely constant 
offset in magnitude $\Delta m$.  Consider constant flat models with 
$w=-1, 0.7$.  By $z=2$ (5;10) their $\wf=-0.08, -0.132$ (-0.011,-0.036; 
-0.002,-0.010) yet their $\Delta m=0.15$ (0.13, 0.11), i.e.~the offset 
is neither negligible, as it would be if only $\wf(z)$ was important, 
nor strictly constant, as it would be if the region where $\Delta\wf(z) 
\approx 0$ gave no contribution to $\Delta m$.  Rather the $\Delta\wf(z) 
\approx 0$ region adds a constant contribution to each model's distance 
$r_l$ which leads to a steady dilution of $\Delta m\sim \log (r_l/r_l')$. 


Inertia is also responsible 
for the turnover occurring at higher redshifts than the $\wf$ crossing. 
While this cannot be made rigorously quantitative with a solely intuitive, 
rather than numerical, approach the $\wf$ analysis does provide a good 
rough estimate (and a lower limit) for the redshift range critical for 
distinguishing the influence of dark energy in the Hubble diagram. 

One can see how well this approximation works in the numerical results 
of Figure 5, for both constant and evolving equations of state. 
While $\wf$ predicts turnovers at $z=(0.8,0.5,0.35)$ and (0.96,0.6,0.34) 
in the $w_0=-1$ and $-0.8$ cases respectively for increasing $w_1$, they 
occur at $z=(1.7,0.9,0.6)$ and ($>$2,1.13,0.55).  (Note that asking 
where the $w$'s rather than $\wf$'s cross is more accurate because 
dilution acts opposite to inertia, but it is less physically justified and 
is undefined in the case of constant $w$ models.)  

\midinsert 
\centerline{
\epsfxsize=3.5truein
\epsfbox{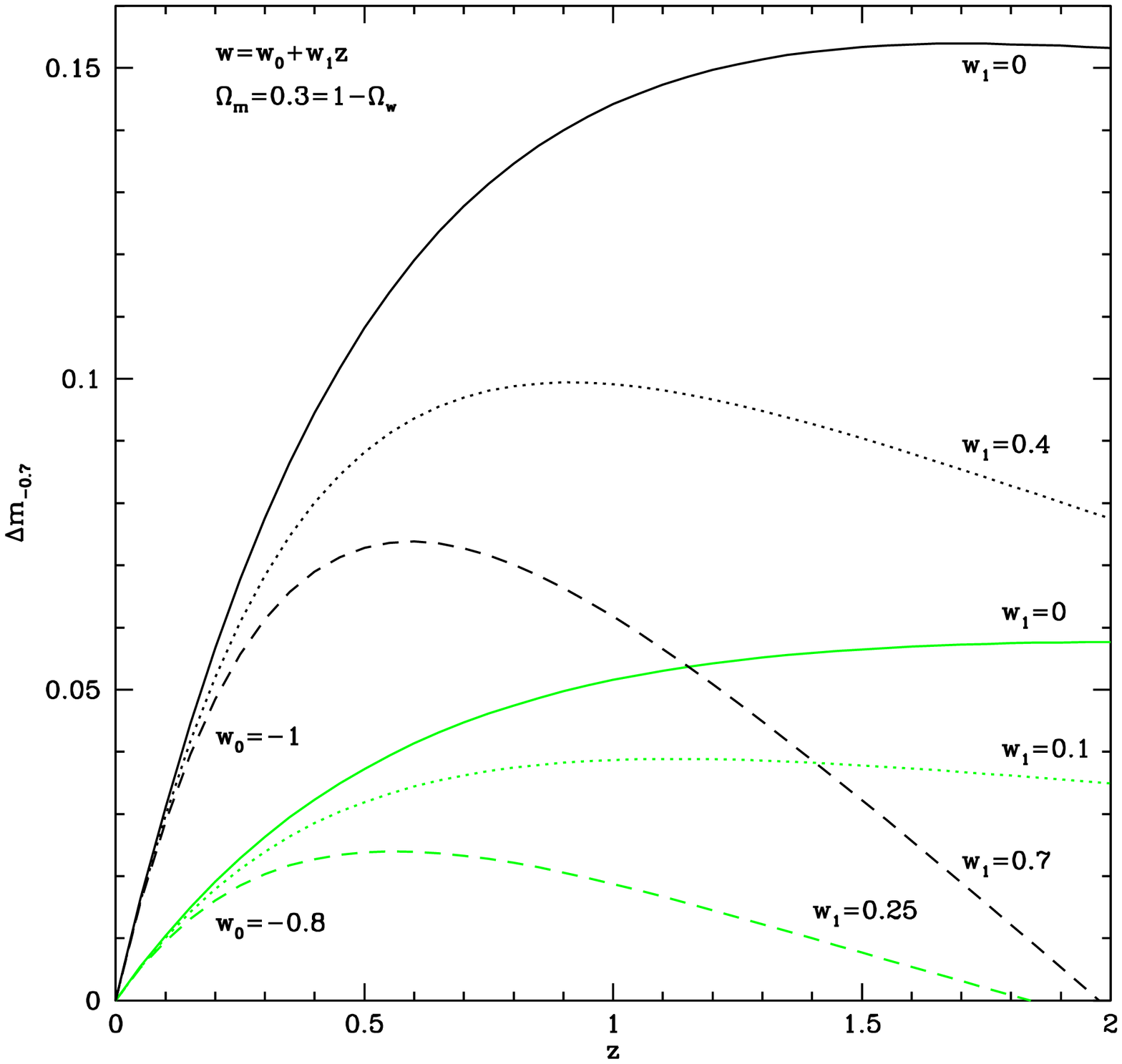}
} 
\nobreak\noindent 
FIG.~5.\quad Differential magnitude-redshift relations are plotted for 
flat, linearly evolving dark energy models $w=w_0+w_1z$ relative to the 
constant model $w=-0.7$.  Prospective SNAP (SuperNova/Acceleration Probe) 
error bars of 0.02 in magnitude 
will be able to distinguish between constant and evolving dark energy 
and also between sufficiently different evolution behaviors by observing 
in redshift out to $z=1.7$. 
\endinsert 

Two interesting points 
are that for all six models the $\wf$ criterion defines a redshift 
about half of the redshift at peak magnitude deviation (turnover) and the 
magnitude deviation there is consistently within 11-13\% of the peak 
deviation, due to the gradualness of the leveling or turnover.  If these 
characteristics were shown to 
be fairly general (as they seem to be given the wide range of $w_0$ and 
$w_1$ in the six models) then the $\wf$ criterion could in fact be 
calibrated to be a fairly robust predictor of both the turnover redshift 
and magnitude deviation. 

\bip 

\leftline{\bf Summary} 
\mip 

The Hubble diagram of Type Ia supernovae is well suited to probing the 
cosmological model for the existence and properties of dark energy.  Such 
a component is expected to dominate the energy density and dynamics, most 
evident in the acceleration of the expansion, at redshifts of $z\le 
0.5-1$.  Thus at these moderate redshifts these models can clearly be 
distinguished from pure matter, or other positive energy, models.  However 
to discriminate {\it between} dark energy models, whether between  
cosmological constant, constant $w$ tracking models, or 
evolving quintessence models, it is necessary to extend the survey depth 
to $z\approx 1-2$.  This reveals 1) the physical imprint of differential 
acceleration, 2) the dilution of the effective equation of state due to 
the increasing dominance of matter, and 3) the turnover to a decelerating 
expansion including the ``inertia'' of the magnitude-distance relation that 
integrates over the equation of state from the source epoch to the present. 

This is shown, along with a cautionary note, in Figure 6.  Although all 
the models graphed have their ``action'' redshifts $z_{eq}$ and $z_{ac}$ 
at $z<0.7$, observations extending only to $z=0.7$ are clearly 
insufficient to probe the cosmological model.  One could not tell 
whether one is dealing with a constant equation of state different from a 
cosmological constant, a rapidly evolving dark energy model, or a purely 
cosmological constant model at a slightly different energy density. 
Surveys extending out to $z\approx 1.5-2$ can make such distinctions, 
even down to fairly fine differences (the curves were chosen 
to represent roughly those ``confusion limits'').  One caution however is 
that complementary constraints, e.g.~on $\omm$, $\omt$ or the low redshift 
behavior, from other cosmological probes or low redshift supernovae data, 
play a crucial role in limiting the parameter space of alternate models. 

\midinsert 
\vskip-0.2truein 
\centerline{
\epsfxsize=3.5truein
\epsfbox{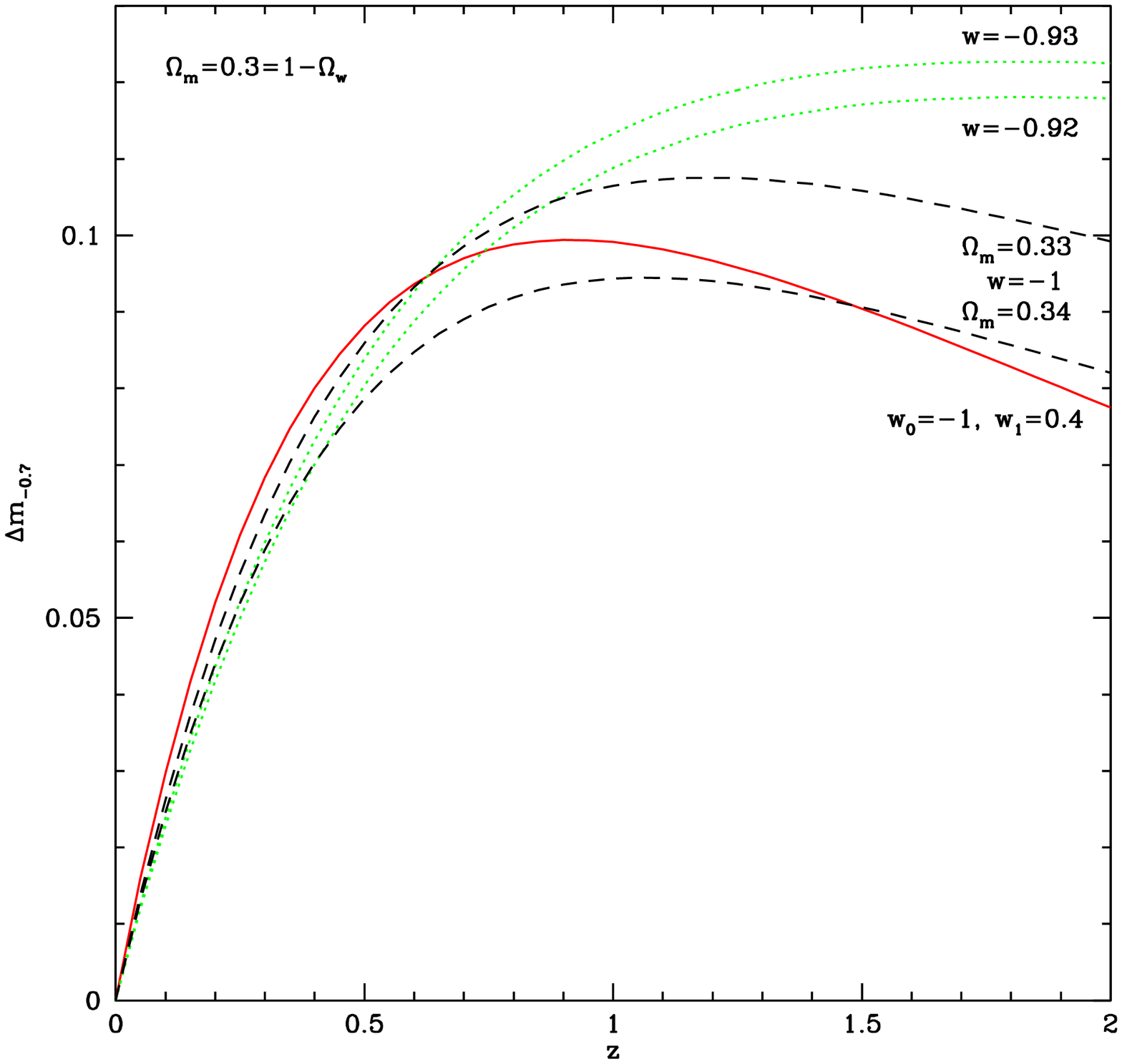}
} 
\nobreak\noindent 
FIG.~6.\quad A ``confusion plot'' of Hubble diagrams is shown with nearly 
degenerate low redshift behavior.  Although at $z>0.7$ all these models are 
matter dominated and decelerating, it is only in this redshift region that 
this cosmological probe becomes useful, able to distinguish different 
dark energy equations of state and constant from evolving $w$.  Note 
however that large uncertainty in other variables such as $\omm$ can 
erode the model parameter determination. 
\endinsert

For the parameter ranges of interest the general rules of thumb in finding 
the ``sweet spot'' of the optimum redshift depth for discrimination between 
models, especially distinguishing between evolving $w(z)$ and a constant 
$w_c$ model, are that the redshift range contributions are: 

\bu significant out to $z$ such that $w(z)\approx w_c$,  
i.e.~similar $\wf$'s 

\qquad (as the magnitude curves diverge); 

\bu mild at higher $z$ as $\wf$'s become similar 

\qquad (differential curve levels off but redshift baseline increases); 

\bu negligible at higher $z$ if the $\wf$'s strongly cross 

\qquad (turnover in differential magnitude; note dilution ensures 
the $\wf$'s never diverge 

\qquad sufficiently at high $z$ to give a large magnitude difference 
of the reversed sign). 

\nobreak\noindent 
Due to the deceleration dilution effect therefore, for the models 
considered here the most likely optimal depth for a SNAP supernovae 
survey is $z\approx 1.5-2$. 

\vfil\eject\bye